# Cybersecurity Software Tool Evaluation Using a 'Perfect' Network Model

Jeremy Straub
Institute for Cyber Security Education and Research
North Dakota State University
1320 Albrecht Blvd., Room 258
Fargo, ND 58108
Phone: +1-701-231-8196
Fax: +1-701-231-8255
Email: jeremy.straub@ndsu.edu

**Abstract**

Cybersecurity software tool evaluation is difficult due to the inherently adversarial nature of the field. A penetration testing (or offensive) tool must be tested against a viable defensive adversary and a defensive tool must, similarly, be tested against a viable offensive adversary. Characterizing the tool's performance inherently depends on the quality of the adversary, which can vary from test to test. This paper proposes the use of a 'perfect' network, representing computing systems, a network and the attack pathways through it as a methodology to use for testing cybersecurity decision-making tools. This facilitates testing by providing a known and consistent standard for comparison. It also allows testing to include researcher-selected levels of error, noise and uncertainty to evaluate cybersecurity tools under these experimental conditions.

**Introduction**

Cybersecurity is, inherently, a battle between attackers and defenders. Attackers attempt to break into systems while defenders attempt to prevent this from happening. In some cases, defenders seek to test their network by having others try to break in to it to identify potential problems before they can be exploited by actual attackers with nefarious purposes. These surrogate attackers are commonly referred to as penetration testers.

Problematically, characterizing an organization or system's level of security, objectively, is quite difficult. A penetration test may find limited vulnerabilities due to the system being very secure. However, tester inexperience (or inexperience with a particular type of system or infrastructure) can also create a similar result. Similarly, attackers may be very successful against organizations and systems with limited defensive capabilities. Succeeding at an attack, foiling an attacker or failing to is not a precise indication of the capabilities of cybersecurity staff. Developing techniques for characterizing cybersecurity staff capabilities, beyond tests of knowledge and skill demonstrations common to certification testing, is an ongoing area of research.

Approaches based on live system testing [1], hybrids of simulation and live system testing [2] and simulation [3-4] have been proposed. The concept of "Stackelberg planning" [5] pits attackers against defenders in a simulation system. Li, Yan and Naili [6] proposed using artificial intelligence techniques to find "optimal attack path[s]". Whole-of-system [7-8] and more focused approaches have been proposed.

Both attackers and defenders benefit from the use of tools that automate numerous tasks, ranging from scanning to identifying vulnerabilities to protect or exploit. These tools can save time of attackers and defenders alike.  They can also increase the thoroughness of cybersecurity activities. They can help defenders and penetration testers scan the entire network for issues to correct. They can also help attackers scan and exploit vulnerabilities. On both sides, tools that go beyond basic scanning are readily available. Attackers (and penetration testers) benefit from attack automation tools [9] while defensive staff benefit from systems that can automatically keep computers up to date with patches and automate the detection [10] of and response to [11] a potential attack. Like with human attackers and defenders, effectively assessing the efficacy of tools is inherently difficult.

To evaluate the software for operations and network attack results review (SONARR), a tool that is designed to perform simulated penetration testing on a digital twin or cousin of systems that are life, safety or mission-critical [12], a new testing technique was developed. This technique utilizes simulated information technology (IT), operational technology (OT) and functionality technology (FT) networks to provide a testing ground and characterize the performance of a cybersecurity decision-making system.

**Method Details**

The proposed method builds on a method proposed for machine learning applications [13] that utilized a 'perfect' rule-fact expert system network as the basis for evaluating the performance of other AI techniques.  This section begins with a discussion of this technique, which provides the foundation for the method proposed herein.  Next, the system that is used to make and evaluate the 'perfect' network is briefly reviewed.  Finally, the proposed method is presented and discussed.

*'Perfect' Expert System for Evaluating Machine Learning Systems*

The 'perfect' expert system network served as the reality of the tests that other techniques attempted to learn about and predict.  To evaluate an AI technique, inputs and outputs were supplied to the expert system and those inputs and outputs provided the basis for training and evaluating the systems being tested.  The evaluation metric was the accuracy of the system being tested's outputs as compared to the outputs of the 'perfect' expert system.  Figure 1 depicts the use of the 'perfect' expert system network for testing a neural network.

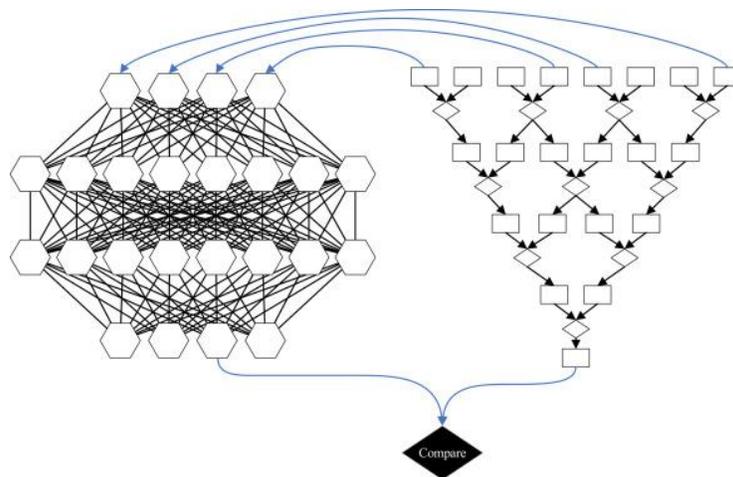

**Figure 1.** Use of a 'perfect' expert system for training and evaluating a neural network [13].

*SONARR and Network Models*

The proposed 'perfect' network for cybersecurity system evaluation is based on the SONARR software [12] and the paradigm for penetration testing systems that cannot be subjected to a penetration test (P2SCP) [14].  While originally designed for modeling systems that cannot be tested using manual and automated techniques which interact with the live systems, SONARR is also useful for a variety of other uses, including managing and prioritizing manual and automated penetration testing.  SONARR creates a network that serves as a 'digital twin' off the system being tested.  Computers and networking equipment are modeled as objects called containers within the software and link objects interconnect these containers representing network interconnections.  Notably, containers and links are versatile and can also be used to represent other phenomena of interest to models and analysis, including human activities such as social engineering.

Both links and containers have properties called facts which store information about the devices which is relevant to determining whether they may be susceptible to different attack techniques.  Some properties that have a shared meaning across different devices or links within the network are associated with common property objects to embody this association.  They still have unique values that can be changed and are stored separately; however, they have a shared meaning that can be used during network evaluation.  Global facts can also be used to represent environmental conditions and to store other information that isn't associated with a particular computer, networking device or interconnection link between them.

The final critical component of the SONARR system is rules.  Rules are used to model attack techniques and, thus, to determine whether an attacker is able to reach and compromise systems on a network.  Rules can model both technology-based attacks (such as a denial of service attack or permissions escalation) as well as other attacks (such as the use of a social engineering technique or phishing).  Rules are associated with links and consider an origin container, destination container and the link when being evaluated.  The origin and destination containers, as well as the link, must all match the requisite pre-conditions for the rule to be run.  If a rule is run, the post-condition values are then used to set fact values on the origin and destination nodes as well as to set the values of global facts that are not associated with a particular container or link.

*'Perfect' Network for Cybersecurity System Evaluation*

The goal of the 'perfect' network for cybersecurity system evaluation is to quantify the accuracy of the system being tested.  For offensive and penetration testing systems, this means determining how frequently the system selects the same attack techniques and paths as the attacker, thereby demonstrating that the system has inferred the attacker's decision-making framework.  For defensive systems, this means quantifying how frequently the system effectively prevents and/or responds to an attacker's activities to block or mitigate an attack.  A specific methodology for each type of testing (offensive / penetration testing and defensive) is presented in the following subsections.

*Use for Testing Offensive Cybersecurity / Penetration Testing Decision-Making Tools*

This section describes the use of the proposed methodology for testing offensive cybersecurity and penetration testing decision making tools.  While the two uses have inherently different goals, the underlying processes used by them are similar, as penetration testing seeks to predict the prospective actions of an adversary to allow defenders to prospectively prevent them. Thus, the models created using this methodology are particular well suited for this predictive task.

Fundamentally, the proposed methodology seeks to evaluate the efficacy of an analysis or decision-making system at identifying and emulating the decision-making processes of an adversary or competitor in planning and executing an attack through a network. This analysis can be applied to the pre-planning of complete attacks or to evaluate adaptive decision making, which is effectively a planning process with new or altered information being considered.

The evaluative metric that is utilized is the level of similarity between the attack path used by the 'perfect' network and decision-making framework, which represents the reality that the tool is trying to learn and model, and the attack path generated by the tool and its models. This similarity metric considers both the path through the network (i.e., what computers or devices are attacked and in what order) as well as the techniques that are used. The weighting between the two is inherently subjective and will need to be determined on a case-by-case basis, in light of research and analysis goals.

To illustrate this, Figures 2 and 3 show models, generated by the SONARR software, of two different attack pathways that have been identified through the same network. This visualization shows the devices and connections utilized (through highlighting of nodes and bolding of lines, respectively) and the number of times a device was involved in a path (through the level of shading). It does not, though, depict the order of the attack through the network or the techniques utilized for each attack step. Both of these are also integral to analysis.

Notably, high-level comparisons could be made considering only the nodes and links utilized (and, prospectively, the frequency of particular nodes and links' use in paths). This would allow more detailed analysis to only be focused on attack paths that have a given level of similarity, saving time.

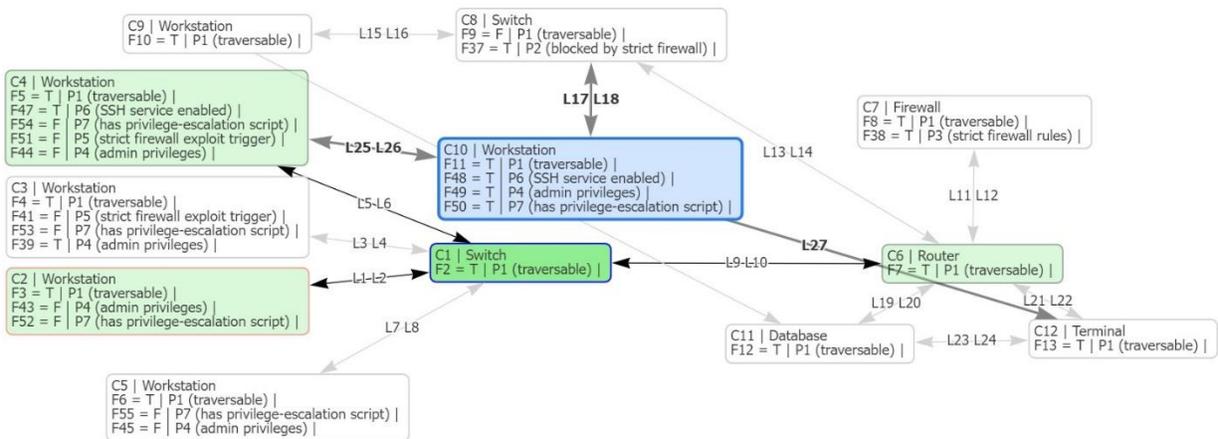

**Figure 2.** Example network with a shorter path chosen.

The process utilized for this analysis, which is presented in Figure 4, begins with the creation of a test network. This is a model of the devices and connections on a given network, which will be utilized for tool evaluation. The properties of devices and connections that are relevant to determining the success or failure of an attack should be included in this model. Notably, modeling can be done at a user-desired level of fidelity; however, the results will be similarly precise or imprecise (i.e., potentially generating false positives or negative results due to a lack of detail). The exact implementation of the model that is used for testing is tool dependent. Modeling in SONARR is described, in detail, in [12,14-16].

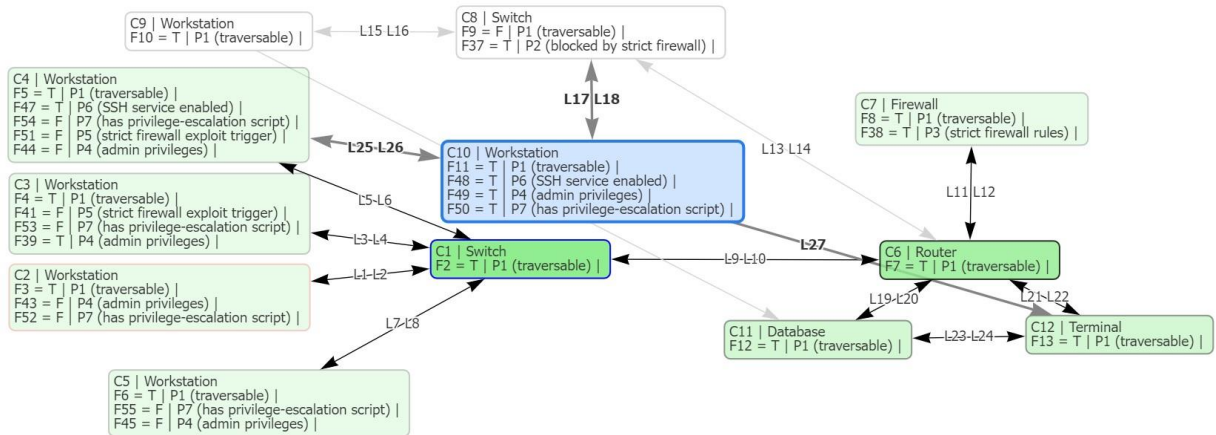

**Figure 3.** Example network with a longer path chosen.

The next step is to create a ruleset which represents the attacks available for attackers to use. In SONARR, these are implemented as rules with pre- and postconditions that model what characteristics devices and their interconnection need to have for an attack to be successful (preconditions) and what the results of an attack are (postconditions).

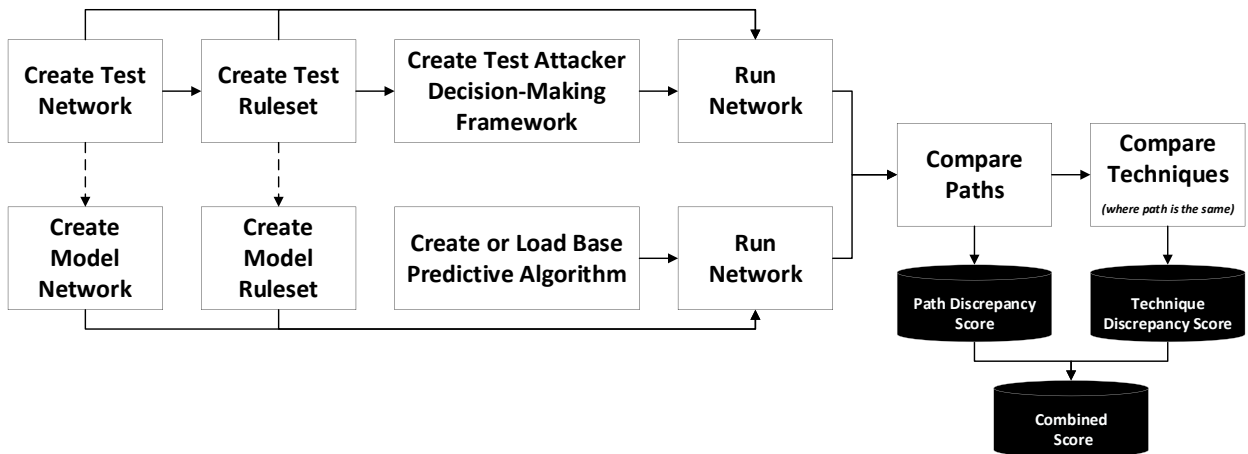

**Figure 4.** Algorithm for testing offensive cybersecurity / penetration testing decision-making systems.

Following this, a decision-making model for the 'perfect' attacker (i.e., the attacker that the system being tested is attempting to predict the decisions of) needs to be created. For a SONARR-style network, this requires the identification of goodness metrics for an attack that the attack decision-making seeks to optimize and the relative weightings applied to these metrics.

The system being tested must also be prepared. The exact particulars of how to do this would inherently vary by the software used. The example in Figure 4 presumes that a SONARR-style 'perfect' network is being used to evaluate the performance of another instance of SONARR in regards to analyzing and determining the decision-making of the 'perfect' attacker. Thus, a model network and model ruleset are created, as they would be for a real-world network, by analysts. This step can involve introducing errors, noise or 'fog of war' to model the difficulties related to real world data collection and to characterize tools' performance under known (and controlled) levels of these characteristics.

The system being tested must also be initialized with a predictive algorithm to use initially.  The normal method of selecting or creating this algorithm, for the system being tested, should be used.  The system being tested will, presumably, seek to refine this algorithm during the learning and testing process.

Once all of the requisite components are in place, both the 'perfect' model/tool and the model/tool being tested are run and the results are compared.  This comparison should consider both the path taken and the techniques that are used as part of this path.  As previously mentioned, analysis of the devices and interconnections in an attack path may be useful to prevent wasting resources on performing detailed analysis of very dissimilar attack paths.  Similarly, attack techniques will typically only be compared when attack paths use the same devices and interconnections.  Some analysis, though, may find value in comparing the techniques used, at a high level, prior to (or without) considering the paths taken.

While the description here has discussed pre-planning of an attack path, the method can also be readily used for testing real-time decision-making in addition to pre-planning decision-making.  Real-time decision making can be treated as a planning problem with new or updated information supplied.  Thus, this process can be used for each time-step in operations and the results (similarities of path and techniques used) can be compared over a collection of timesteps, instead of making only a single comparison.

*Use for Testing Defensive Cybersecurity Decision-Making Tools*

Focus now turns to the use of an adaptation of the proposed method for evaluating defensive cybersecurity tools, processes and their associated paradigms.  This is necessarily more complex, as it requires modeling the adversarial or competitive interaction between the offensive and defensive players (which can be humans, automated or a combination of the two) to evaluate the efficacy of defensive decision-making.

The proposed method for defensive tool/process evaluation, which is shown in Figure 5, uses a similar process for the offensive side.  A 'perfect' testing network, attack ruleset and decision-making framework are created, as was described in the process for evaluating offensive / penetration testing tools.  However, the system being modeled and what happens during the run process are inherently different.  Additionally, an arbiter process is used to facilitate testing and data collection.  The arbiter uses the 'perfect' model network that is created as the basis for its operations.

The proposed method considers both proactive and reactive defensive activities.  Proactive activities are performed once the model network is known, but before attacks start.  Reactive activities are the detection and mitigation activities that are taken in response to an attack.  Some tools may only implement one style of activity; however, the methodology supports both.

Once applicable rulesets (for techniques) and decision-making frameworks (to determine what metrics to optimize for, for each phase) are available, the simulation run process commences.  Notably, this process is inherently iterative.  First, all proactive techniques are performed.  Then, the simulation loop commences.  During each iteration, the offensive simulation process launches an attack and advises the arbiter of this.  The arbiter uses its model and ruleset to determine what symptoms and indications of the attack would exist and sends these to the defensive process. The defensive process then advises the arbiter of its response. This cycle repeats. When it is complete, the defensive tool and/or paradigm is evaluated based on its level of success in preventing and responding to attacks. The basis for this success (i.e., particular preventative or response techniques) can also be identified and analyzed.

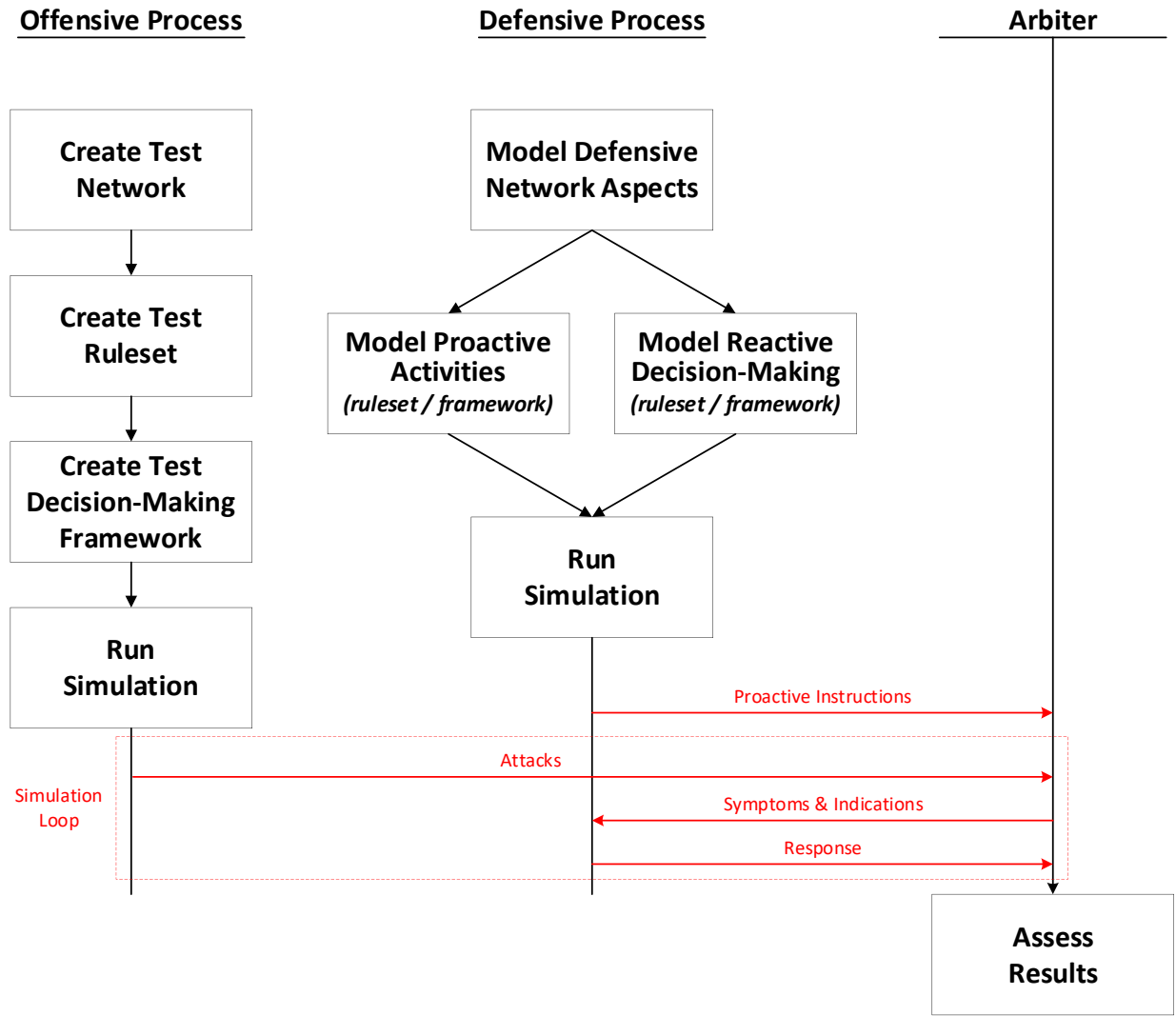

**Figure 5.** Algorithm for testing defensive cybersecurity / penetration testing decision-making systems.

While this basic approach may be suitable for many applications, there are a number of potential augmentative steps, beyond those shown in Figure 5. These augmentations can facilitate a more nuanced analysis, but also add complexity to the process (and associated analysis).

First, the basic methodology assumes that the attack process has an accurate model of the network being attacked. This is not a valid assumption, in most real-world scenarios. Thus, the 'perfect' model that is initially created could be used only by the arbiter process and the attack process could be given a degraded model with error, noise or ambiguity introduced. This may reduce the repeatability of testing; however, it also facilitates introducing the realism of potentially apparently-erratic attack decisions which are caused by these factors.

Second, the base methodology assumes that the attack process will carry out a pre-planned attack without considering the proactive or reactive measures implemented by the defensive process.  While this may be an accurate representation of some attack (particularly smaller attacks that occur within a limited duration of time), many attacks will involve adaptation to current conditions.  Thus, to consider this, the model network used by the attack process could be updated (either with completely accurate information or with degraded information, as discussed above) after the proactive tasks are complete and/or after each iteration of attack and defense.

Finally, modeling constraints may be valuable for some research.  These could prospectively be introduced throughout the simulation.  For example, the defensive process could be allowed to run only a given number of proactive processes (or given a time cap for these processes, if the time required was specified for each technique).  Similarly, responsive actions could be given a limited amount of time to be performed (again, based on techniques specifying the time required) before additional attacks are run.

*Expanded Uses*

In addition to the uses for testing offensive / penetration testing and defensive tools, this method could prospectively be expanded for other uses. For example, it could be modified to facilitate assessing intrusion detection systems or systems that analyze and draw conclusions from computer forensics data.

Any system or process that relates to, uses or analyzes data from, or supports or impacts cybersecurity attacks and defensive activities could prospectively be tested using an augmentation of the proposed method.  The method would provide data and support analysis related to the attack and defense decision-making and would, for most applications, need to be augmented with an application-specific framework, process or methodology extension.

Beyond this, this methodology could be expanded to support analysis related to many of the other broad collection of uses that have been identified for the SONARR software.  These include cybersecurity topic areas such as social engineering, supply chain threats, attack detection and insider threat detection.  More generally, the method may have application to other competitive or adversarial processes.  This would need to be evaluated more specifically on an application area-by-application area basis.

**Summary**

This paper has presented a methodology for testing tools and techniques related to the cybersecurity attack and defense process.  It has been designed around the SONARR software; however, it could prospectively be implemented with other similar software or without the use of software altogether.  The proposed method uses a 'perfect' model of a cyberattack/penetration test and/or cyberdefense activities to evaluate tools that attempt to characterize them or make decisions related to attacks, penetration testing and defense.

While the focus of this methodology is on testing and characterizing tools and techniques, these are, inherently, the first steps of generating data that can support answering broader cybersecurity research questions.  Thus, by enabling this data collection and analysis, the proposed methodology facilitates research in a wide variety of areas related to the competitive / adversarial research domain of cybersecurity. It may also have application to other research areas that study phenomena with competitive and adversarial characteristics.

## Acknowledgements

Thanks is given to Matthew Tassava, Cameron Kolodjski, Jordan Milbrath and other team members for their work on the SONARR software.